\documentclass{epl}

\title
{Exact results of the ground state and excitation properties of
a two-component interacting Bose system}
\author{You-Quan Li\inst{1}, Shi-Jian Gu\inst{1},
Zu-Jian Ying\inst{1}, and Ulrich Eckern\inst{2}}
\institute{ \inst{1}Zhejiang Institute of Modern Physics, Zhejiang University, Hangzhou 310027, P.R. China\\
\inst{2}Institute of Physics, University of Augsburg, 86135
Augsburg, Germany}


\begin{document}

\maketitle

\begin{abstract}
We study a one-dimensional Bose system with repulsive
$\delta$-function interaction in the presence of an SU(2) intrinsic
degree of freedom on the basis of the coordinate Bethe-ansatz. The
ground state and the low-lying excitations are determined by both
numerical and analytical methods. It is shown that the ground
state is an isospin-ferromagnetic state, and the excitations are
composed of three elementary particles: holons, antiholons, and
isospinons. The isospinon is a triplet coupled to the
``ferromagnetic'' background anti-parallelly.
\end{abstract}

\pagestyle{myheadings} \markright{You-Quan Li {\it
et al.}: Two-component interacting bosons}

 Exactly solvable models
\cite{WiegmannAP83,AndreiEMP83,Haldane83,deVegaNP85,%
KorepinBook,LiebW,LiebL,Gaudin67,Yang67,Sutherland68,%
ChoyHaldane,Schlottman,Gaudin71,Schulz,Woynarovich85,LiG98}
play an important role in physics, specifically in the investigation of
one-dimensional (1D) interacting many-particle systems. They
have served as a source of inspiration for the
understanding of non-perturbative phenomena in correlated
electronic systems; e.~g., the spinon was explicitly
characterised on the basis of the exact solution of the
Hubbard model \cite{LiebW}. Among these models, an earlier prototypical
one is the model of 1D bosons with repulsive $\delta$-function
interaction, which was solved \cite{LiebL} by means of the Bethe-ansatz.
This method was also applied to solve the problem of spin-1/2
fermions \cite{Gaudin67,Yang67} with $\delta$-function
interaction; in fact, 
Yang \cite{Yang67} already suggested
a general strategy for multi-component systems.
Various extensions include the study of electrons on a crystalline lattice
\cite{LiebW}, the generalization to higher
symmetries \cite{Sutherland68,ChoyHaldane,Schlottman}, and
applications to different boundary conditions
\cite{Gaudin71,Schulz,Woynarovich85,LiG98}. Nevertheless, 
SU(2) bosons with $\delta$-function interaction, as
far as we are aware, have not been studied until now -- in contrast,
obviuously, to the ``spin 1/2'' fermionic case, i.~e.\ a
two-component model with anti-symmetric permutation.
Recently, however, a two-component Bose gas was created in
magnetically trapped $^{87}$Rb by rotating the two hyperfine
states into each other with the help of a slightly detuned Rabi
oscillation field \cite{WilliamsTH,WilliamsEXP}; and it was noticed
\cite{Janson} that the ground state of a Bose system can be
surprisingly different from the scalar Bose system,
once the particles acquire an intrinsic degree of freedom. 

Therefore, in 
order to obtain non-perturbative insight into the features of
one-dimensional SU(2) bosons with repulsive $\delta$-function interaction,
we study a model which
is integrable. Pointing out the connection with the coupled
Gross-Pitaevski equation \cite{Gross,Pitaevski}, we solve the
Bethe-ansatz equations for bosons with an SU(2) intrinsic degree of
freedom. 
The ground state properties and the low-lying excitations are
studied by both, a numerical calculation and in the thermodynamic limit.
Unlike a spin-1/2 Fermi system, the ground state of
the present model is a ``ferromagnetic'' state, consistent
with the result of a variational approach \cite{Janson}. The
charge-isospin phase separation is confirmed, and the isospinon is a
triplet instead of a doublet (the latter being well known for the 
spinon in the spin-1/2
Fermi system).

The two-component Bose gas is known
to satisfy the coupled  Gross-Pitaevski equations:
\[
i\hbar\frac{\partial}{\partial t}
 \pmatrix{\psi_1 \cr \psi_2}=
  \pmatrix{\hat{G}_1 & \hat{P}^* \cr \hat{P} & \hat{G}_2}
   \pmatrix{\psi_1 \cr \psi_2 } ,
\]
where $\hat{G}_b=-\frac{\hbar^2}{2m}\nabla^2_b + V_b({\bf r})
 +\sum_a u_{ab} |\psi_a|^2$, $\hat{P}=\hbar\Omega/2$, and $a,b=1,2$;
$\Omega$ is the Rabi oscillation field frequency, and $V({\bf r})$ 
the trapping
potential.

We consider the isotropic limit, in which the strengths of interaction between
inter-species and intra-species are the same ($u_{ab}=c$); then
the model is integrable. Considering further the system trapped in a
1D ring of length $L$, and introducing
$\phi_1=(\psi_1+\psi_2)/\sqrt{2}$,
$\phi_2=(\psi_1-\psi_2)/\sqrt{2}$, we obtain from the above
equation (for real $\Omega$) the following equivalent
Hamiltonian:
\[
H=\int dx \Bigl[\partial_x\phi_a^*\cdot\partial_x\phi_a
    +c\phi^*_a\phi^*_b\phi_b\phi_a
     -(-1)^a \Omega\phi^*_a\phi_a\Bigr]
\]
with summation over repeated indices implied; natural units are
adopted for simplicity. The fields obey bosonic commutation
relations, $ [\,\phi^*_a(x),\,\phi_b(y)\,]
=\sum_{n\in{Z\!\!\!Z}}\delta_{a b}\delta(x-y-nL). $ The Rabi
oscillation field contributes a Zeemann-type term. To avoid
confusion with conventional spins,
we denote the generators of the
isospin SU(2) by ${\cal I}$; correspondingly $ [{\cal I}^+, {\cal
I}^-]=2{\cal I}^z. $ The intrinsic degree of freedom can be
specified by the eigenvalues of ${\cal I}^z$, or by isospin
up and down.

The Bethe-ansatz equations for two-component bosons
are  given as follows:
\begin{eqnarray}
e^{ik_jL}=-\prod^{N}_{l=1}\Xi_1(k_j-k_l)
   \prod_{\nu=1}^{M}\Xi_{-1/2}(k_j-\lambda_\nu) ,
    \nonumber\\
1=-\prod^{N}_{l=1}
    \Xi_{-1/2}(\lambda_\gamma-k_l)
     \prod^{M}_{\nu=1}
      \Xi_1(\lambda_\gamma-\lambda_\nu) ,
\label{eq:BAE}
\end{eqnarray}
where $\Xi_\beta(x)=(x+i\beta c)/(x-i\beta c)$. Equation (\ref{eq:BAE})
determines the value of the quasi-momenta $\{k_j\}$ and the
isospin rapidities $\{\lambda_\nu\}$ for a $N-2M+1$ fold multiplet
characterised by the total isospin ${\cal I}_{tot}=(N-2M)/2$. These
equations are similar to the Bethe-ansatz equation of 
\cite{Sutherland75}, except for a variation in the exponential.

Equation (\ref{eq:BAE}) is obtained as follows.
Applying the Hamiltonian
to the Hilbert space of $N$ particles, and considering its first
quantized version on the domain ${\sf l\!\!R}\setminus\{{\sf
l\!\!P}_{ij}\}$ where ${\sf l\!\!P}_{ij}:=\{x|x_i-x_j=0\}$ is the
hyperplane defined by the $\delta$-function singularity, we see
that only the $N$-dimensional Laplacian remains in the
Schr\"{o}dinger operator. Thus $N$-dimensional plane waves are
solutions. We sum up all the plane waves with
wave vectors which are are just permutations of a definite  ${\bf
k}=(k_1, k_2,\ldots,k_N)$ according to the Bethe-ansatz strategy.
Integrating the Schr\"{o}dinger equation across the hyperplanes,
we obtain $S(k_i-k_j)=[k_i-k_j-ic{\cal P}]/[k_i-k_j+ic]$, which
connects the wave functions defined on the regions separated by
the hyperplanes, and $\check{S}:={\cal P}S$ (${\cal P}$ stands for
the spinor representation of the permutation group $S_N$) relates
the coefficients of different plane waves in the same region. 
The bosonic permutation symmetry
(instead of the antisymmetry) was imposed when solving for the
$S$-matrix. Analogous to the case of spin-1/2 fermions
\cite{Yang67}, the periodic boundary condition leads to an
eigen-equation for the product of the $S$-matrices. As the
$S$-matrix satisfies the Yang-Baxter equation, the
quantum inverse scattering method \cite{Faddeev} is applicable.
After writing out the fundamental commutation
relations, and evaluating the eigenvalues of the reference state
$|\omega\rangle=|\uparrow\uparrow\ldots\uparrow\rangle$, one 
immediately recognises the differences to the case of spin-1/2 fermions.
For example, $A(\xi)|\omega\rangle=\prod_l
(\xi-\xi_l-ic)/(\xi-\xi_l+ic)|\omega\rangle$, and
$D(\xi)|\omega\rangle=\prod_l
(\xi-\xi_l)/(\xi-\xi_l+ic)|\omega\rangle$, in the notion of
\cite{Faddeev,Kulish}. Consequently, we obtain Eq.
(\ref{eq:BAE}). We note that the Bethe-ansatz
strategy implies the existence of infinitely many constants
of motion, $\sum_j k^n_j=constant$, in addition to the usual
energy $E=\sum_{l=1}^N k_l^2+\Omega(N-2M)$, and momentum
$P=\sum_l^N k_l$.

Taking the logarithm of Eq. (\ref{eq:BAE}) leads to
\begin{eqnarray}
k_j=\frac{2\pi}{L}I_j+\frac{1}{L}\sum^{N}_{l=1}\Theta_1(k_j-k_l)
     &+&\frac{1}{L}\sum_{\nu=1}^{M}\Theta_{-1/2}(k_j-\lambda_\nu),
     \nonumber\\
2\pi J_\gamma=\sum^{N}_{l=1}\Theta_{-1/2}(\lambda_\gamma-k_l)
  &+&\sum^{M}_{\nu=1}\Theta_1(\lambda_\gamma-\lambda_\nu),
\label{eq:logBAE}
\end{eqnarray}
where $\Theta_\beta(x):=-2\tan^{-1}(x/\beta c)$;
both the quantum numbers $I_j$ and $J_\gamma$ take $integer$ or
$half$-$integer$ values, depending on whether $N-M$ is $odd$ or $even$.
For comparison, the Bethe-ansatz equation for
spin-1/2 fermions does not only lack the first summation,
but also has an opposite sign in the second summation
in the first line of Eq. (\ref{eq:logBAE}).
The momentum is easily obtained from Eq. (\ref{eq:logBAE}),
$P=\sum_l k_l=(\sum_l I_l-\sum_\nu J_\nu)2\pi/L$.

It is instructive to analyse Eq. (\ref{eq:logBAE})
in the strong and weak coupling regimes.
For a strong interaction, $c\rightarrow\infty$, the wave function
vanishes for any $x_i=x_j$, and hence the bosons avoid each other
like fermions which is in agreement with the discussion of
quantum degeneracy in trapped 1D gases \cite{Petrov}.
On the other hand, in the weak coupling limit, $c\rightarrow 0$,
using $\Theta_1(x)\rightarrow-\pi{ \rm sgn}(x)$, and
$\Theta_{-1/2}(x)\rightarrow\pi{ \rm sgn}(x)$ for $x\gg 1$, 
Eq. (\ref{eq:logBAE}) becomes 
\begin{eqnarray}
k_j+\frac{\pi}{L}\sum_{l=1}^N{\rm sgn }(k_j-k_l)
   -\frac{\pi}{L}\sum_{\nu=1}^M{\rm sgn }(k_j-\lambda_\nu)
     =\frac{2\pi}{L}I_j,
       \nonumber\\
\sum_{l=1}^N {\rm sgn }(\lambda_\gamma-k_l)
  -\sum_{\nu=1}^M{\rm sgn }(\lambda_\gamma-\lambda_\nu)
    =2J_\gamma.\;\;
\label{eq:weakBAE}
\end{eqnarray}
The subscript of the isospin rapidity $\lambda_\gamma$ can be
chosen in such a way that $J_\gamma$ is arranged in increasing
order; then the second equation of Eq. (\ref{eq:weakBAE}) turns into
\begin{equation}
\sum_{l=1}^N{\rm sgn }(\lambda_\gamma -k_l)
=2J_\gamma+2\gamma-M-1.
\label{eq:2weakBAE}
\end{equation}
Because $|J_\gamma|<(N-M+1)/2$ for a given $M$ and
$M\leq N/2$ due to the restriction given by the Young tableau, the minimum
value of the right hand side of Eq. (\ref{eq:2weakBAE}) is $-N+2$.
This requires that the smallest $k_l$ must be smaller than the smallest
$\lambda_\nu$, otherwise the left hand side would be $-N$.
Equation (\ref{eq:2weakBAE}) also implies
\begin{eqnarray}
\sum_{l=1}^N \Big[ {\rm sgn}(\lambda_{\gamma+1}-k_l)
  -{\rm sgn}(\lambda_\gamma-k_l) \Big] 
=2(J_{\gamma+1}-J_\gamma+1).\;\;\;
\end{eqnarray}
Thus, for $J_{\gamma+1}-J_\gamma=m$, there must exist exactly $m+1$
solutions of $k_l$ satisfying $\lambda_\gamma<k_l<\lambda_{\gamma+1}$.
Furthermore, from the first equation of Eq. (\ref{eq:weakBAE}), we obtain
\begin{eqnarray}
k_{j+1}&-&k_j
  -\frac{\pi}{L}\sum_{\nu=1}^{M}
    \Big[ {\rm sgn}(k_{j+1}-\lambda_\nu)
     -{\rm sgn}(k_j-\lambda_\nu) \Big]
\, = \frac{2\pi}{L}(I_{j+1}-I_j -1).
\label{eq:1weakBAE}
\end{eqnarray}
Obviously, for $I_{j+1}-I_j=n$, there will be
$k_{j+1}-k_j=2n\pi/L$ if there is a $\lambda_\gamma$
such that $k_j<\lambda_\gamma<k_{j+1}$, otherwise
$k_{j+1}-k_j=(n-1)2\pi/L$. Thus an isospin rapidity of value
$\lambda_\mu$ always
repels the quasi-momenta away from that value.
As a result, an existing $\lambda_\mu$ will suppress the density of
states in $k$-space at the point $k=\lambda_\mu$.
The more isospin rapidities there are, the higher the energy
will be.
Thus the ground state of SU(2) interacting bosons is no longer a
SU(2) singlet, but an isospin ``ferromagnetic'' state
which differs from the Fermi case considerably.

\begin{figure}
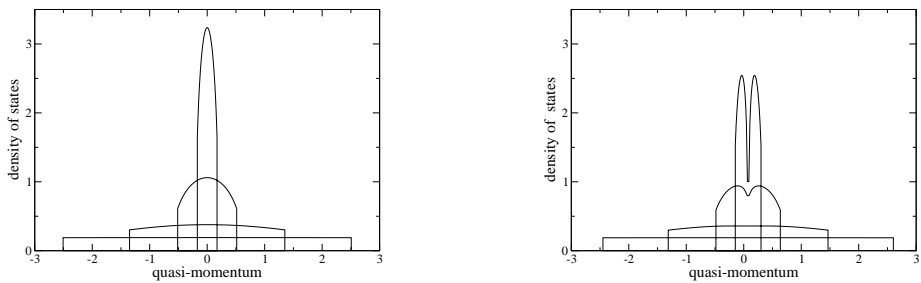

\twoimages[width=5cm]{rhok_g.eps}{rhok_j0.eps} \caption{ The
density of state per length in $k$-space for the ground state
(left) and for the state in the presence of one isospin rapidity
$\lambda$ by choosing $J_1=0$ (right). The distribution gradually changes
from a ``histogram'' to a narrow peak for strong to weak coupling,
$c=$ 10, 1, 0.1, 0.01. The calculation is done for
$L=40$, $N=40$. The left panel is similar to  Fig. 2 of [7].}
\label{fig:density}
\end{figure}

\begin{figure}
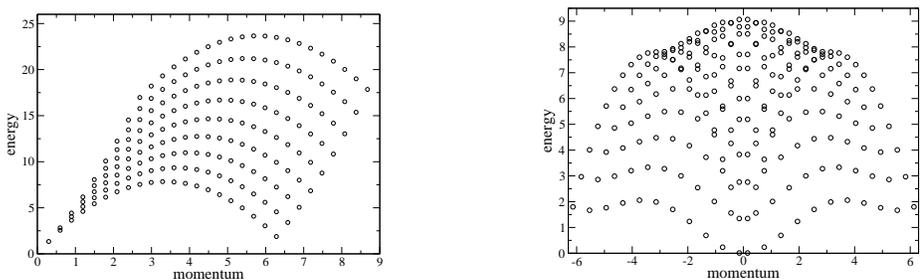

\twoimages[width=5cm]{p-h10.eps}{isp-h10.eps} \caption{The
holon-antiholon excitation spectrum (left) and holon-isospinon
excitation spectrum (right), calculated for $L=20$, $N=40$, and $c=10$.}
\label{fig:spectra}
\end{figure}

For $N$ particles, the ground state is characterised by a one-row
$N$-column Young tableau $[ N ]$, of which the quantum numbers are
$\{ I_j\}:=\{-(N-1)/2,...,(N-1)/2\}$ and $\{J_\gamma\}= empty$.
For this state Eq. (\ref{eq:logBAE}) reduces to the case studied in
\cite{LiebL}, but the ground state of the present model
is an $(N+1)$-fold multiplet with ${\cal
I}^2=N(N+2)/4$. The density of states per length for the ground
state is plotted in Fig. \ref{fig:density} (left) for various coupling
constants. The ``particle''-hole (or maybe to be called holon-antiholon)
excitation is defined by the quantum numbers $I_1 =
-(N-1)/2+\delta_{1,j_1}$ (for $1\leq j_1\leq N$), $I_j =
I_{j-1}+1+\delta_{j,j_1}$ (for $j=2,...,N-1$), and
$|I_N|\geq(N+1)/2$. Figure \ref{fig:spectra} (left) shows the
corresponding excitation spectrum. The isospinon-holon excitation
is characterised by the Young tableau $[N-1, 1]$, i.~e., $M=1$. In
comparison to those of the ground state, the quantum numbers $\{I_j\}$
change from half-integer to integer or vice versa; accordingly,
$I_1 = -N/2+\delta_{1,j_1}$ (for $1\leq j_1\leq N+1$), $I_j =
I_{j-1}+1+\delta_{j,j_1}$ (for $j=2,...,N$), while $J_1=I_1+n$ so
that $I_1<J_1<I_N$. This is an $(N-1)$-fold multiplet with ${\cal
I}^2=N(N-2)/4$. The excitation spectrum is plotted in Fig.
\ref{fig:spectra}. In fact two branches of the quasi-particle excitations
were recently observed in a two-component 
condensate \cite{Goldstein} by
means of techniques from nonlinear optics. The density of states
for $J_1=0$, $j_1=1$ is  plotted in Fig. \ref{fig:density}. In
comparison to the ground state where no isospin rapidity exists, a
rift emerges at the position of the isospin rapidity for
small $c$ that is consistent with our previous analysis for weak
coupling.

In the thermodynamic limit,
the Bethe-ansatz equations lead to the following integral equations
for the density of roots and the density of
holes, respectively, in quasi-momentum
and isospin rapidity spaces:
\begin{eqnarray}
\rho(k)+\rho_h(k)
  &=& \frac{1}{2\pi}+\int_{Q}^{Q}dk'\rho(k')K_1(k-k')
  \, - \int_{-B}^{B}d\lambda'\sigma(\lambda')K_{1/2}(k-\lambda'),
           \nonumber\\
\sigma(\lambda)+ \sigma_h(\lambda)
  &=& \int_{-Q}^{Q}dk'\rho(k')K_{1/2}(\lambda-k')
  \, - \int_{-B}^{B}d\lambda'\sigma(\lambda')K_1(\lambda-\lambda'),
\label{eq:thermodynamics}
\end{eqnarray}
where $K_\mu(x)=\pi^{-1}\mu c/(\mu^2 c^2+x^2)$. The limits of
integration, $Q$ and $B$,
are determined to be consistent with $\int_{-Q}^{Q}\rho(k)dk=N/L$, and
$\int_{-B}^{B}\sigma(\lambda)d\lambda=M/L$. It is easy to check by
Fourier transformation that the state with $B=\infty$ and $\sigma_h=0$ is
an isospin singlet, but it is not
the ground state. The ground state corresponds to
$\sigma=\rho_h=0$ in Eq. (\ref{eq:thermodynamics}), i.~e.\
it is an isospin ``ferromagnetic''
state, in agreement with the result of mean field
theory \cite{Janson}. The two-particle case is a pedagogical example:
For the two-body Schr\"odinger equation in the
center-of-mass frame, the permutation of particle coordinates
equals the parity reflection of their relative coordinate. The
oscillation theorem in quantum mechanics tells that the spatial
wave function without nodes, an even parity solution, yields
the lowest energy. If it possesses an SU(2) intrinsic degree of
freedom, the intrinsic wave function must be symmetric
(anti-symmetric) to keep the total wave function with the lowest
energy being symmetric (anti-symmetric). Then the ground state of
the Bose system (Fermi system) is of ``ferromagnetic''
(``anti-ferromagnetic'') character.

The highly degenerate ferromagnetic ground state, obtained in the case of a
vanishing Zeemann term, will split up into Zeemann sublevels
once the external field is applied.
The ground state hence becomes a polarized state
once the Rabi field, which breaks the SU(2) symmetry, is turned on.

In order to evaluate the excitation energies we put
$\rho(k)=\rho_0(k)+\rho_1(k)/L$ ($\rho_0$ refers to the ground state).
In the presence of the isospin degree of freedom, there will 
be a holon-isospinon
excitation, in addition to the holon-antiholon
excitation. The latter is created by a hole inside
the quasi Fermi sea $\bar{k}\in[-k_F, k_F]$, and an additional
$k_p$ outside it, i.~e.
\[
\rho_1(k)+\delta(k-\bar{k})
  =\int_{-k_F}^{k_F}dk'\rho_1 K_1(k-k')+K_1(k-k_p).
\]
The excitation energy consists of two terms:
$\Delta E=\int k^2\rho dk +k_p^2=\varepsilon_h(\bar{k})+\varepsilon_a(k_p)$,
where the holon energy $\varepsilon_h$, and the antiholon energy
$\varepsilon_a(k_p)=-\varepsilon_h(k_p)$, are given by
\begin{eqnarray}
\varepsilon_h(y)&=&-y^2+\int_{-k_F}^{k_F}k^2\rho^h_1(k,y)dk,
   \nonumber\\
\rho^h_1(k,y)&+&K_1(k-y)=\int_{-k_F}^{k_F}dk'K_1(k-k')\rho^h_1(k',y).
\label{eq:holon}
\end{eqnarray}
Flipping one isospin corresponds to adding one isospin rapidity to the
background of the ferromagnetic ground state,
which inevitably brings about one hole
in the $k$-sector.
The excitation energy $\Delta E =\int k^2\rho_1 dk$ is obtained from
\[
\rho_1(k)+\delta(k-\bar{k})=\int_{-k_F}^{k_F}dk'K_1(k-k')\rho_1(k')
  -K_{1/2}(k-\lambda);
\]
consequently, $\Delta E=\varepsilon_h(\bar{k})+\varepsilon_{i}(\lambda)$.
Here $\varepsilon_h$ is given by Eq. (\ref{eq:holon}), 
and $\varepsilon_{i}$ by
$\varepsilon_{i}(\lambda)=\int k^2\rho^{i}_1(k,\lambda)dk$, with
\[
\rho^{i}_1(k,\lambda)+K_{1/2}(k-\lambda)
 =\int_{-k_F}^{k_F}dk'K_1(k-k')\rho^{i}_1(k',\lambda).
\]

\begin{figure}
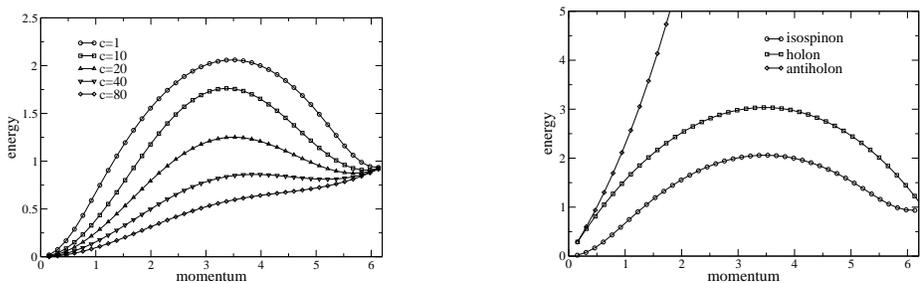

\twoimages[width=5cm]
 {isospinon.eps}{excitons.eps}
\caption{Dispersion relations of the isospinon for
different coupling constants (left); the curves from top to bottom
correspond to $c=$ 1, 10, 20, 40, and 80, respectively. Right: 
dispersion of 
antiholon, holon, and isospinon, for $c=1$
($L=20$,  $N=40$).}
\label{fig:dispersins}
\end{figure}

In conclusion we found three elementary
quasi-particles: holon, antiholon and isospinon. From the
asymptotic behaviour of those basic modes, for $\bar{k}$, $k_p$,
and $\lambda$ tending to $k_F$, we find that both the holon-antiholon
and holon-isospinon excitations are gapless. The related
dispersions for finite $N$ are plotted in Fig.
\ref{fig:dispersins}. Different from the spinons in a Fermi
system, the isospinon here is a triplet which always couples to
the ``ferromagnetic'' background anti-parallelly. Although it is
always accompanied by a charge excitation (holon) for single or odd
number of isospinons, the isospinons can be excited in pairs
without exciting the U(1) charge mode. Because of the coupling
between the charge sector and the isospin sector in
Eq. (\ref{eq:thermodynamics}), both cases bring about changes in
the quasi-momentum distribution and hence lead to an
excitation energy.
The charge-isospin phase separation predicted by mean field theory
\cite{phaseMFT} is clearly confirmed in the present case due to the
structure of Eq. (\ref{eq:thermodynamics}). The holon and antiholon
are quasi-particles created in momentum space, while the
isospinon behaves like a dark soliton \cite{Carr} in the isospin
sector that tends to decrease the total
isospin ${\cal I}$ eigenvalue by one. 

Considering the experiment on a two-component Bose gas whose
transverse excitations are frozen out, such that 
the dynamics becomes essentially one-dimensional \cite{Ketterle}, it
is expected that the above results
can be confirmed through a careful measurement 
of the excitation spectra, which in particular should show the 
isospin excitations. However, the question has to remain open
of how much can be learned from our model -- which is an integrable one --
about the phenomenon of Bose-Einstein condensation, as e.~g.\ the inclusion
of a realistic trapping potential into the model should be important.

\medskip
\centerline{***}

This work was supported by the trans-century project of the Ministry of
Education and the NSFC for distinguished youth. YQL acknowledges
the support of the AvH-Stiftung, and  thanks J. Cardy, J. Carmelo,
H. Frahm, F.D.M. Haldane, T.C. Ho, M. Oshikawa, and H. Saleur for
useful discussions; also thanks to F. Guinea for interesting remarks
on the manuscript. Partial support from the Deutsche Forschungsgemeinschaft
(SFB 484, SPP 1073) is acknowledged.

\newpage

\end{document}